\documentclass[runningheads]{llncs}
\usepackage[T1]{fontenc}
\usepackage{todonotes}
\usepackage{graphicx}
\usepackage{booktabs}
\usepackage{xcolor}
\usepackage{comment}
\usepackage{booktabs}
\usepackage{tabularx}
\usepackage{amsmath}
\usepackage{amssymb}  

\newcommand{\perfect}{\checkmark}
\newcommand{\underfit}{$\sim$}
\newcommand{\impossible}{$\times$}
\newcommand{\overfit}{$+$}

\definecolor{placeholdercolor}{RGB}{255,80,80}

\begin{document}
\title{Towards a faithful stochastic model for brain digital twins}
\titlerunning{Towards a faithful stochastic model for brain digital twins}
\author{Abdelhamid Rouatbi\inst{1}\orcidID{0009-0007-1617-6827} \and
Alexandre Muzy\inst{2}\orcidID{0000-0002-8594-1672} \and
Eugene Syriani\inst{1}\orcidID{0000-0001-6527-1651}}
\authorrunning{A. Rouatbi et al.}
\institute{
DIRO, Universit\'e de Montr\'eal, Montr\'eal, Canada\\
\email{abdelhamid.rouatbi@umontreal.ca}, \email{syriani@iro.umontreal.ca}
\and
ILLS, CNRS, McGill University, ETS, MILA, Universit\'e Paris-Saclay, Montr\'eal, Canada\\
\email{alexandre.muzy@cnrs.fr}
}
\maketitle
\begin{abstract}
Twinning the brain means reproducing its electrical activity with fidelity. This activity arises from many neuronal mechanisms across several scales, from a single neuron to whole brain regions. Current large initiatives rely on models that each capture only a few of these mechanisms. Some act at the scale of a single neuron, others at the scale of neuronal populations. No single model captures both scales with fidelity. We propose a new stochastic model, based on a combination of Random Neural Networks and Markov-Modulated Poisson Processes, that would improve the fidelity of a brain digital twin. We map the model to the Discrete Event System Specification. This yields an initial executable simulation model suitable for incorporation into a brain digital twin.


\end{abstract}

\section{Introduction}
A digital twin is a virtual replica of a real system, built to mirror it with high fidelity~\cite{Rasheed2020}. Achieving such fidelity requires models of the mechanisms that govern the system. In complex systems, these mechanisms are numerous and heterogeneous. This multiplicity makes modeling difficult. The brain is an extreme case: understanding brain function through electrical activity is notoriously difficult~\cite{mascart2022scalability}. Neuronal mechanisms unfold across several scales, from the dynamics of a single neuron to the collective behavior of entire brain regions.

Brain digital twins aim to capture these mechanisms with fidelity~\cite{muzy2026braintwins}. Several large initiatives pursue this goal, as illustrated by the detailed Blue Brain reconstruction~\cite{markram2015reconstruction} and the MICrONS project, which mapped cortical structure and function~\cite{microns2025functional}. These efforts rely on models of electrical activity that each capture only a subset of the relevant mechanisms. At the scale of a single neuron, dedicated models describe the membrane potential~\cite{hodgkin1952quantitative}, action potential emission~\cite{izhikevich2003simple}, bursting~\cite{hindmarsh1984model}, and the refractory period. At the scale of neuronal populations, neural mass models average this activity, describing connectivity~\cite{sanzleon2015mathematical} and firing synchrony~\cite{wilson1972excitatory}.  

For brain digital twins, a central challenge is to identify modeling abstractions that adequately represent relevant neuronal mechanisms while remaining computationally tractable at large scales. Since brain electrical activity is sparse in both time and space, discrete-event models provide a suitable framework for its simulation~\cite{mascart2022scalability}. In addition, neuronal activity exhibits intrinsic variability: even under identical stimulation, a neuron or neuronal network may produce different responses~\cite{azouz1999cellular}. Stochastic models can account for this variability while abstracting the underlying microscopic mechanisms, thereby providing a statistically meaningful representation of neuronal dynamics. 

Existing stochastic models provide different abstractions of neuronal activity, each capturing particular neuronal mechanisms. Rather than relying on a single existing model to represent all relevant mechanisms, we propose to combine complementary mechanisms from different stochastic modeling approaches into a new model. In particular, Random Neural Networks (RNNs) provide an explicit representation of network connectivity and excitatory and inhibitory interactions, while Markov-modulated Poisson processes (MMPPs) capture temporal variations in firing activity~\cite{gelenbe1994gnet,fischer1993mmpp}. Combining these complementary mechanisms provides the basis for a stochastic model capable of representing neuronal dynamics at both the individual-neuron and network levels.

In this work, we provide a comparative review of existing stochastic discrete-event neuron models, assessing the neuronal mechanisms they represent and their limitations with respect to biologically relevant firing and interaction dynamics. We propose the \textit{Markov-Modulated Random Neural Network} (MM-RNN), a novel stochastic neuron model that combines the network-level interaction mechanisms of RNNs with Markov-modulated firing dynamics. We provide an approach that enables specifying high-level parameters to automatically generate the technical neurons and network parameters into a Discrete Event System Specification (DEVS) simulation model~\cite{zeigler2018theory,hans,muzy2007designing}. This model represents neuronal mechanisms consistently at the neuron level, composes them into networks through modular coupling, and provides an executable model that can be directly simulated.

This work provides a basis for studying how neuronal mechanisms interact to produce collective electrical activity and for incorporating the model into a simulation service of brain digital twins. Scalability remains achievable as our approach is compatible with Mascart et al's work on the scalability of large neural network simulations~\cite{mascart2022scalability}.

\section{Background}

This section defines fundamental neuronal mechanisms and reviews how stochastic neuron models represent them.

\subsection{Fundamental neuronal mechanisms}
Biological neurons communicate through electrical signals whose generation and propagation depend on mechanisms operating at both the individual-neuron and network levels. We consider six fundamental mechanisms that characterize neuronal electrical activity and provide the basis for evaluating the stochastic models presented in the following section.

\subsubsection{Membrane potential}

The membrane potential represents the electrical state of a neuron and arises from the potential difference across its membrane~\cite{hodgkin1952quantitative}. It evolves over time in response to the intrinsic dynamics of the neuron and to incoming synaptic stimuli. Excitatory inputs tend to move the membrane potential toward the threshold for spike generation, whereas inhibitory inputs move it away. The membrane potential therefore provides an internal state through which incoming activity is integrated and neuronal excitability is determined.

\subsubsection{Spike emission}

Neurons communicate by generating action potentials, or spikes. A spike is produced when the electrical state of the neuron reaches the conditions required for firing~\cite{izhikevich2003simple}. Once emitted, it propagates along the axon and can influence postsynaptic neurons through synaptic connections. At an abstract level, neuronal activity can therefore be represented as a sequence of discrete spike events characterized by quantities such as firing rates, inter-spike intervals, and spike-time distributions. A stochastic model must consequently specify the process governing spike occurrence.

\subsubsection{Refractory period}

Following the emission of a spike, a neuron undergoes a refractory period during which its ability to generate another spike is temporarily reduced. During the absolute refractory period, another action potential cannot be generated, whereas during the relative refractory period firing remains possible but requires stronger stimulation. This mechanism constrains the temporal structure of spike trains by limiting how closely successive emissions can occur and therefore bounds the maximum firing rate.

\subsubsection{Bursting}

Under certain conditions, neurons exhibit bursting behavior, consisting of periods of rapid spike emission separated by intervals of lower activity or silence~\cite{hindmarsh1984model}. Bursting introduces temporal structure beyond individual spike events. Representing this mechanism requires firing dynamics that can change over time, allowing the neuron to alternate between regimes associated with different levels of activity.

\subsubsection{Synaptic connectivity}

Neurons interact through synaptic connections that transmit emitted spikes from presynaptic neurons to postsynaptic targets. Depending on the synapse, the interaction may be excitatory, increasing postsynaptic excitability, or inhibitory, decreasing it. The strength of this influence is determined by the synaptic connection. Connectivity therefore defines both the structure of the neuronal network and the pathways through which electrical activity propagates. Its representation requires specifying which neurons interact, the type of interaction, and its relative strength.

\subsubsection{Synchrony}

Neuronal activity can also exhibit collective temporal organization. Synchrony refers to the coordinated firing of multiple neurons within a sufficiently narrow temporal interval~\cite{wilson1972excitatory}. It characterizes temporal dependencies between spike trains rather than treating neuronal activity independently. Such coordinated activity contributes to collective neuronal dynamics and is also central to Hebbian descriptions of neuronal assemblies, where correlated activity is associated with the strengthening of connections between neurons~\cite{hebb1949organization}. Representing synchrony therefore requires capturing dependencies between firing events across neurons.

Together, these mechanisms describe complementary aspects of neuronal electrical activity. Membrane potential, spike emission, refractory behavior, and bursting primarily characterize individual-neuron dynamics, whereas synaptic connectivity and synchrony describe interactions and collective behavior at the network level. The stochastic models considered in the following section differ in how they represent these mechanisms.

\subsection{Stochastic neuron models}

Several classical stochastic discrete-event models can be used to represent neuronal spiking activity. These models differ in their representation of time, internal neuronal state, spike generation, and interactions between neurons. Consequently, each model captures a different subset of the neuronal mechanisms introduced in the previous section.

\subsubsection{Bernoulli process}

The Bernoulli process provides a simple discrete-time model of stochastic event generation. Time is divided into fixed intervals, and at each time step an event occurs independently with a fixed probability. When applied to neuronal activity, an event represents the emission of a spike~\cite{galves2015modeling}. Therefore, the probability of emission determines the average firing activity of the neuron, while the intervals between successive spikes are determined by the sequence of Bernoulli trials.

This representation provides a simple stochastic abstraction of spike emission but does not explicitly represent the internal state of the neuron. In particular, the emission probability is fixed and does not depend on a membrane potential or on previously received stimuli. Consequently, mechanisms such as refractory periods and bursting cannot be represented directly. At the population level, correlated activity can nevertheless be introduced by defining joint emission probabilities across multiple neurons, allowing synchronous spike emissions to be represented~\cite{grun2002unitary}. The model remains inherently discrete in time, and its temporal resolution is determined by the selected time step.

\subsubsection{Poisson process}

The Poisson process provides a continuous-time alternative for modeling stochastic spike emission. In a homogeneous Poisson process, events occur independently at a constant average rate, while the time intervals between successive events are exponentially distributed. Interpreting these events as spikes yields a neuron that fires stochastically at a prescribed mean firing rate~\cite{urdapilleta2009firing}. Unlike the Bernoulli process, spike times are defined in continuous time.

The homogeneous Poisson process provides a concise representation of irregular spike trains but retains no internal neuronal state. Its firing rate remains constant and is independent of incoming stimuli and previous emissions. It therefore does not directly represent mechanisms such as membrane potential dynamics, refractory periods, or changes between different regimes of activity. At the population level, correlated spike trains can be constructed using approaches such as Generalized Thinning and Shift (GTaS)~\cite{trousdale2013generative}. A common parent process generates events that are assigned to selected subsets of neurons, possibly with temporal shifts, thereby introducing controlled synchronous and correlated activity between their spike trains.

\subsubsection{Markov chains}

Markov chains introduce an explicit discrete internal state and can therefore represent richer neuronal dynamics. A neuron is described by a finite set of states and probabilistic transitions between them. In a neuronal interpretation, particular states can be associated with spike emission, such that a spike is produced when the model transitions into a firing state~\cite{nossenson2010modeling}. Depending on the formulation, transitions may occur at discrete time steps or in continuous time.

The state-based representation allows different phases of neuronal activity to be modeled explicitly. For example, a refractory period can be represented by requiring the neuron to enter one or more refractory states following an emission before returning to a state from which another spike can occur. Similarly, burst-like behavior can be represented through sequences of states that favor several closely spaced emissions. This additional expressiveness comes at the cost of a more elaborate state space, since each additional neuronal behavior must be encoded through appropriate states and transitions. Moreover, spike emission is associated with transitions between abstract states rather than represented as a distinct stochastic process, and the basic Markov-chain formulation does not explicitly represent synaptic connectivity or interactions between neurons.

\subsubsection{Markov-modulated Poisson process}

An MMPP combines a continuous-time Markov chain with a Poisson process~\cite{fischer1993mmpp}. The Markov chain represents a set of activity regimes, with each state associated with a different Poisson event rate. As the Markov chain changes state, the event rate changes accordingly. The resulting process therefore retains the event-based representation of the Poisson process while introducing temporal variations in its intensity.

For neuronal modeling, the generated events represent spikes and the state of the underlying Markov chain determines the current firing regime of the neuron. States associated with low firing rates can represent periods of low activity, whereas states associated with higher rates can represent periods of increased activity. Transitions between these regimes allow the model to represent temporally varying firing patterns and, in particular, burst-like activity. As with the Poisson model, synchrony across neurons can also be introduced through correlated point-process constructions such as GTaS.

The MMPP nevertheless remains primarily a model of spike generation. The state of its modulating Markov chain represents a firing regime rather than the membrane potential of the neuron, and incoming excitatory or inhibitory spikes are not explicitly represented. Consequently, the model does not by itself provide a representation of synaptic connectivity or network interactions.

\subsubsection{Random Neural Networks}

RNNs represent neuronal networks using queueing networks~\cite{gelenbe1994gnet}. Each neuron is modeled as a queue, and its occupancy represents its level of excitation. Neurons receive two types of signals. An excitatory signal increases the excitation level by one, whereas an inhibitory signal decreases it by one when the neuron is excited. A neuron whose excitation level is positive is active and can emit spikes similarly to a Poisson process. Each emission decreases its excitation level and may subsequently be routed to another neuron or leave the network.

This queueing interpretation provides an explicit representation of interactions between neurons. An emitted spike can be routed as an excitatory or inhibitory signal, while routing probabilities determine the connectivity between neurons and the nature of their interactions. These probabilities therefore provide an abstraction of synaptic connectivity and weights. In contrast to the preceding point-process models, neuronal activity consequently depends on the signals received from the rest of the network rather than solely on an autonomous spike-generation process.

The RNN representation nevertheless introduces several abstractions. The excitation level is an integer-valued quantity rather than a direct representation of the continuous membrane potential, and a neuron becomes inactive only when its excitation level reaches zero. Refractory behavior is therefore implicit rather than represented by an explicit refractory state or duration. Moreover, in the standard RNN, firing rates and routing probabilities are static model parameters. Extensions of the RNN introduce synchronized interactions, in which the firing of one neuron can trigger the firing of another toward the same target, thereby providing a mechanism for representing synchronous neuronal activity~\cite{gelenbe2008random}.

\subsubsection{Markov-DEVS}

Markov-DEVS specializes the DEVS formalism for the representation of stochastic systems with semi-Markov dynamics~\cite{seo2018devs}. A model evolves between discrete states according to probabilistic transitions, while the time spent in each state can follow general probability distributions. Unlike a conventional Markov or semi-Markov process, Markov-DEVS also provides explicit input and output events and inherits the coupling mechanisms of DEVS, allowing multiple models to be composed hierarchically.

Applied to neuronal modeling, states can represent different neuronal conditions, including firing and refractory states, while transitions determine the stochastic evolution between these conditions. Explicit input events allow the state of a neuron to be affected by spikes received from other neurons, and output events can represent emitted spikes. Through DEVS coupling, individual neuron models can subsequently be connected to form neuronal networks.

Markov-DEVS is therefore sufficiently expressive to represent a broad range of neuronal mechanisms. However, this expressiveness is provided through general-purpose state, transition, input, output, and coupling constructs rather than abstractions specific to neuronal activity. Mechanisms such as membrane potential, spike emission, refractory behavior, and synaptic interactions must consequently be defined explicitly by the modeler. Markov-DEVS thus provides considerable modeling flexibility, but at the cost of greater specification complexity for neuronal models.

\subsection{Stochastic neuron modeling limitations}

\begin{table}[t]
\centering
\setlength{\tabcolsep}{8pt}
\renewcommand{\arraystretch}{1}
\caption{Comparison of models according to their ability to represent different neuronal mechanisms:
\perfect~adequate expressiveness, \underfit~under-expressiveness, \impossible~not expressible, \overfit~over-expressiveness.
}
\label{tab:comparaison-modeles}
\resizebox{\textwidth}{!}{
\begin{tabular}{lcccccc}
\toprule
\textbf{Mechanism}
& \textbf{Bernoulli}
& \textbf{Poisson}
& \textbf{Markov Chain}
& \textbf{MMPP}
& \textbf{RNN}
& \textbf{Markov-DEVS} \\
\midrule
\textbf{Emission}
& \underfit & \perfect & \underfit & \perfect & \underfit & \underfit \\
\textbf{Refractory period}
& \impossible & \impossible & \perfect & \impossible & \underfit & \perfect \\
\textbf{Bursting}
& \impossible & \impossible & \underfit & \perfect & \underfit & \overfit \\
\textbf{Potential}
& \impossible & \impossible & \impossible & \impossible & \underfit & \overfit \\
\textbf{Connectivity}
& \underfit & \underfit & \underfit & \underfit & \perfect & \overfit \\
\textbf{Synchrony}
& \perfect & \perfect & \impossible & \perfect & \perfect & \overfit \\
\bottomrule
\end{tabular}
}
\end{table}

The comparison in Table~\ref{tab:comparaison-modeles} shows that no single model provides an adequate stochastic representation of all neuronal mechanisms considered. Each model represents some mechanisms adequately while exhibiting limitations on others. For example, the Poisson process is well suited to spike emission, but it cannot represent a refractory period or an internal potential. Conversely, Markov-DEVS can express a wide range of mechanisms, but its generality comes with substantial modeling overhead.

Among the models considered, RNNs and MMPPs are closest to the identified needs. RNNs are particularly well suited to network-level modeling, as they explicitly represent excitatory and inhibitory interactions and the connectivity structure between neurons. However, standard RNN routing sends each emitted customer to a single destination, whereas biological spikes are transmitted to all postsynaptic targets. Spike emission therefore remains under-expressed. This limitation also affects refractory periods, which are represented only implicitly: after firing, a neuron can remain inactive only if it receives no excitation. MMPPs, in contrast, support changes in activity regime and burst representation, but they do not directly model interactions between neurons or a membrane potential. A more adequate model could therefore combine both approaches: an RNN whose service rate varies over time according to an underlying Markov chain would preserve the network structure of RNNs while introducing variable activity as in MMPPs.


\section{The proposed model}

In this section, we present a new stochastic discrete-event model that better captures the biological mechanisms discussed above.

\subsection{Motivation}

As discussed in the previous section, existing stochastic neuron models capture different subsets of neuronal mechanisms. Among the models considered, RNNs provide the most comprehensive representation of network-level interactions, including excitatory and inhibitory connectivity, synaptic weights, and synchronized interactions. However, their firing dynamics are governed by static service rates, which limits their ability to represent temporal variations in neuronal activity. In contrast, MMPPs naturally represent changes between different firing regimes, including periods of low and high activity, but do not explicitly represent neuronal interactions or network connectivity. These complementary properties motivate the introduction of the \textit{Markov-Modulated Random Neural Network} (MM-RNN), a new stochastic neuronal model that extends the RNN with Markov-modulated firing dynamics. Each neuron retains the queueing structure and interaction mechanisms of an RNN, while its service rate varies over time according to the state of an associated continuous-time Markov chain.

\subsection{MM-RNN model}

Each neuron $i$ is modeled as a queue whose length $K_i(t)$ represents the excitation level.
External excitatory and inhibitory spikes arrive at each neuron according to independent Poisson processes with rates $\lambda_i^{+}$ and $\lambda_i^{-}$, respectively, increasing or decreasing the neuron's excitation level $K_i(t)$ by one.
A neuron is active whenever its excitation level is strictly positive, $K_i(t) > 0$, and emits spikes according to a Poisson process with the current firing rate $r_i$.
When the excitation level is zero, $K_i(t)=0$, the neuron is inactive and cannot emit spikes.
Upon firing, a neuron generates an event, and its excitation level $K_i(t)$ is decremented.

The outcome of the event is selected probabilistically according to the model routing parameters $P^{+}=(p_{ij}^{+})_{1\leq i,j\leq N}$, $P^{-}=(p_{ij}^{-})_{1\leq i,j\leq N}$, and $Q=(q_{ijk})_{1\leq i,j,k\leq N}$. A spike may:

\begin{itemize}
    \item Be transmitted as an excitatory signal to neuron $j$ with probability $p_{ij}^{+}$.
    \item Be transmitted as an inhibitory signal to neuron $j$ with probability $p_{ij}^{-}$.
    \item Trigger a synchronous interaction involving neurons $j$ and $k$ with probability $q_{ijk}$.
    \item Leave the network.
\end{itemize}

Synchronous interactions $q_{ijk}$ represent second-order effects in which the firing of a first neuron $i$ triggers the firing of a second neuron $j$ toward the same target neuron $k$.
In this case, the target neuron $k$ receives only one unit of excitation.
If the second neuron $j$ is inactive, only the first neuron $i$ loses an excitation unit, with no effect on the target.

The service rate of each neuron $i$ is governed by an associated continuous-time Markov chain $M_i$, referred to as a modulator.
Each state $Z_i(t)$ of the modulator corresponds to a distinct firing rate $r_i(Z_i(t))$, while transitions between modulator states, governed by the generator matrix $G_i$, dynamically alter the neuron's firing behavior.
Thus, the service rate of neuron $i$ at time $t$ is given by $r_i(t)=r_i(Z_i(t))$.

The proposed model addresses all six mechanisms in our comparison, with different levels of expressiveness.
Neuronal emission is represented directly, as active neurons  can generate spikes.
However, spike emission follows a one-to-one interaction pattern, which remains a simplification of biological spike propagation. The refractory period is captured only implicitly by the excitation mechanism: a neuron with zero excitation, $K_i(t)=0$, becomes inactive and cannot emit spikes until it receives a new excitatory input.
The model can represent bursting behavior through Markov-modulated firing rates, which allow the neuronal activity level to vary over time.
However, because neurons are modeled as queues, the effective intensity of bursts is constrained by the accumulation of excitation from incoming spikes.
The neuron's membrane potential is represented by the queue length.
Connectivity is represented through the routing probabilities, which define how spikes are transmitted between neurons.
Finally, synchronous behavior is supported through second-order interaction probabilities.

\subsection{DEVS simulation}

Discrete-event modeling approaches are well suited to spiking neuronal systems.
Neuronal activity is expressed through spike emissions, which occur sparsely in time.
Discrete-event simulation can therefore avoid unnecessary computation between spike events.
We rely on the DEVS formalism as a framework for the hierarchical, modular modeling and simulation of discrete-event systems~\cite{zeigler2018theory}.
A DEVS atomic model encapsulates a component's autonomous behavior.
It maintains an internal state $S$ and defines how long it remains in that state through the time-advance function $ta$.
Before an internal transition, an output may be produced through the output function $\lambda$.
The state $S$ is updated either autonomously through the internal transition function $\delta_{\mathrm{int}}$, or in response to an external input through the external transition function $\delta_{\mathrm{ext}}$.
Coupled models are composed by connecting the input and output ports of atomic models or other coupled models.
When internal and external events occur simultaneously, the confluent transition function $\delta_{\mathrm{conf}}$ defines how these concurrent events are handled.
When multiple components are scheduled to perform an internal transition at the same simulation time, the $\operatorname{select}$ function  determines which component is processed first.

\subsection{Mapping the model onto DEVS}
\label{sec:mapping}

\begin{figure}
  \centering
  \includegraphics[width=\textwidth]{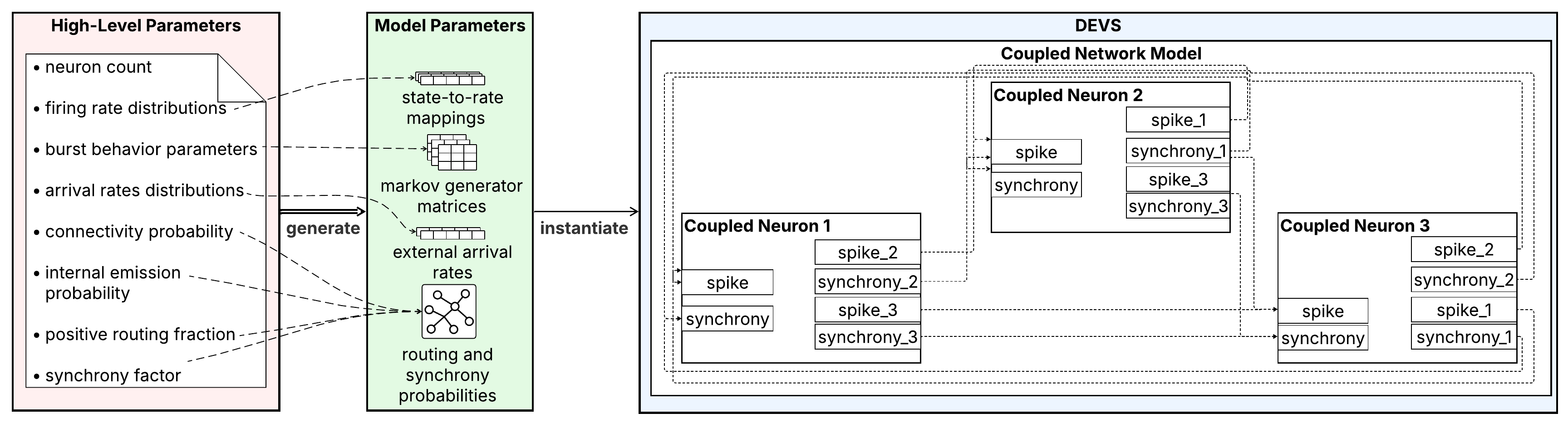}
  \caption{The MM-RNN generation pipeline, from high-level parameters to a coupled DEVS model.}
  \label{fig:conceptual-framework}
\end{figure}

The MM-RNN model is specified by a few intuitive, high-level parameters of the population and its neurons (Figure~\ref{fig:conceptual-framework}). These include the network size $N$; a baseline firing-rate distribution $\mathcal{D}_{r}$; a bursting firing-rate distribution $\mathcal{D}_{r_b}$; the burst frequency $\nu_b$ and average burst duration $\tau_b$; distributions of external excitatory and inhibitory arrival rates, $\mathcal{D}_{\lambda^+}$ and $\mathcal{D}_{\lambda^-}$; the connectivity probability $c$; the proportion of excitatory connections $\alpha$; the internal routing probability $\rho$; and a global synchrony factor $p_s$.

Network connectivity is generated as an Erd\H{o}s--R\'enyi graph,
where each possible connection between two neurons is independently present
with probability $c$. The parameter $\rho$ specifies the total probability
that an emitted spike is routed through an ordinary internal connection,
while $\alpha$ specifies the fraction of this probability assigned to
positive routing.
Accordingly,
$\sum_j p_{ij}^{+}=\alpha\rho$ and
$\sum_j p_{ij}^{-}=(1-\alpha)\rho$.
The global synchrony factor $p_s$ specifies the probability that an emitted
spike participates in a synchronized interaction. The remaining probability
$d = 1-\rho-p_s$ corresponds to the spike leaving the network.

From these parameters, the technical parameters of the markov modulator (MM) and RNN submodels are generated. For each neuron $i$, the baseline and bursting firing rates are sampled from $\mathcal{D}_{r}$ and $\mathcal{D}_{r_b}$, respectively, yielding the firing-rate parameters $r_i$ associated with the states of the Markov modulator $M_i$. Likewise, the external positive and negative arrival rates, $\lambda_i^{+}$ and $\lambda_i^{-}$, are sampled from $\mathcal{D}_{\lambda^+}$ and $\mathcal{D}_{\lambda^-}$. The burst frequency $\nu_b$ and average burst duration $\tau_b$ determine the transition rates of the two-state Markov modulator $M_i$. Network connectivity is generated according to an Erd\H{o}s--R\'enyi graph, which will be used to instantiate the connections between the coupled neuron models ports. For each neuron, the internal routing probability $\rho$ is distributed randomly among its outgoing connections to generate the entries of the routing matrices $P^+$ and $P^-$, with a fraction $\alpha$ assigned to positive routing and $1-\alpha$ to negative routing, such that $\sum_j p_{ij}^{+}=\alpha\rho$ and $\sum_j p_{ij}^{-}=(1-\alpha)\rho$. The synchrony factor $p_s$ determines the probability that a firing event is synchronized. As a simplification, the two participating neurons $j$ and $k$ are selected randomly from the population. The remaining probability $d_i = 1-\rho-p_s$ corresponds to spikes leaving the network.

\begin{figure}
  \centering
  \includegraphics[width=0.8\textwidth]{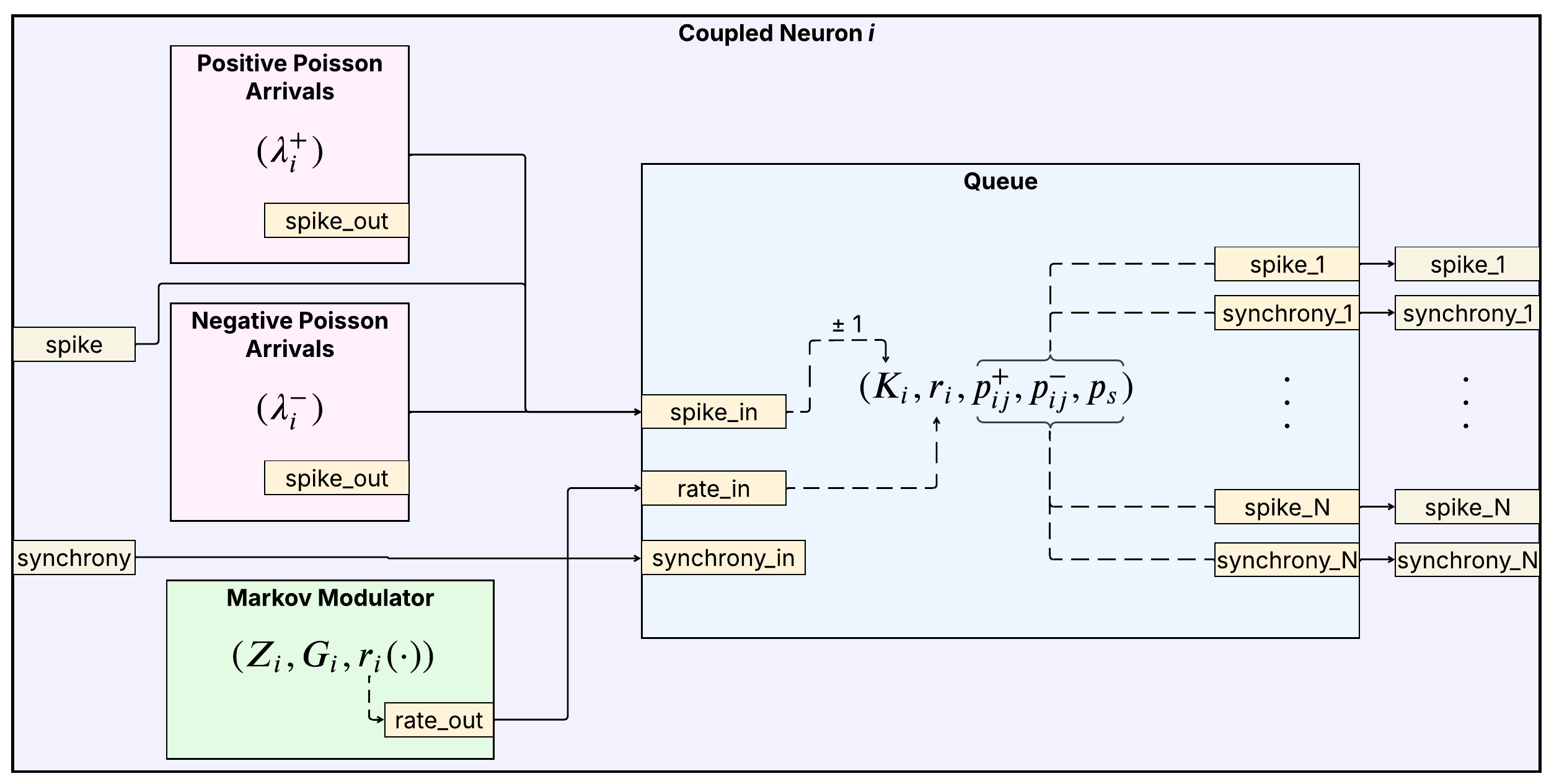}
  \caption{Coupled neuron DEVS model overview.}
  \label{fig:neuron}
\end{figure}

Figure \ref{fig:neuron} details the DEVS realization of a single MM-RNN neuron. Each coupled neuron contains two Poisson source models representing external excitatory and inhibitory arrivals, a Markov modulator governing the current firing rate, and a queue model representing the neuronal excitation state and spike dynamics. The queue additionally receives routed spike and synchrony events from other neurons and emits spike or synchrony events toward the corresponding target neurons.

The \textit{Markov chain model} operates autonomously. Its time-advance
function $ta$ samples the sojourn time in the current Markov state according
to the corresponding exponential distribution. Upon expiration, the output
function $\lambda$ emits the selected successor state to the queue model, and
the internal transition function $\delta_{\mathrm{int}}$ updates the current
Markov state and selects its next successor state. Since the Markov chain
evolves independently of incoming events, no external transition function
$\delta_{\mathrm{ext}}$ is defined.

External arrivals are generated by dedicated \textit{Poisson source atomic
models}. Each source represents an independent excitatory or inhibitory
arrival process. For a source associated with neuron $i$, the time-advance
function samples from an exponential distribution using the corresponding rate, depending on the type of source. Upon expiration, the output function
$\lambda$ emits the corresponding stimulus event to the queue model, and
$\delta_{\mathrm{int}}$ schedules the next arrival.

The \textit{queue model} handles spike dynamics, routing, and changes in the
firing regime. For neuron $i$, when its queue is nonempty, $ta$ samples the
time until the next firing from an exponential distribution with the firing
rate associated with the current Markov state; when the queue is empty,
$ta=\infty$. Upon expiration, the output function $\lambda$ determines the
type of firing event according to the routing parameters $P^{+}$, $P^{-}$,
$Q$, and $d_i$, and emits the corresponding event. For a spike emission event, the target neuron is selected according to $P^{+}$ or $P^{-}$. For a
synchronized event, the target neurons $j$ and $k$ are selected randomly from
the population and a synchrony event is emitted.
The internal transition function $\delta_{\mathrm{int}}$ decrements the queue
length to account for the completed service.

The external transition function $\delta_{\mathrm{ext}}$ handles three types
of input events. Positive and negative stimulus events, whether generated by
external Poisson sources or routed from other neurons, respectively increment
or decrement the queue length. Markov-state updates change the firing rate
used by $ta$. Finally, a synchrony event received  causes $\delta_{\mathrm{ext}}$ to place the queue model in an immediate-firing state when its queue is nonempty.

All autonomously scheduled event times in the MM-RNN are sampled from
continuous distributions. Consequently, the probability that an independently
scheduled internal event coincides exactly with an external event, or that two components are scheduled to realize an internal transition at the same time is zero. Therefore, confluent transition functions $\delta_{\mathrm{conf}}$ and $\operatorname{select}$ functions for coupled models are not
required.

\subsection{Initial evaluation}
We implemented an initial prototype of the MM-RNN as a coupled DEVS model in Python\-PDEVS~\cite{PyPDEVS}, following the mapping described in Section~\ref{sec:mapping}, and simulated individual and small networks of neurons.
\begin{figure}[h]
  \centering
  \includegraphics[width=0.8\linewidth]{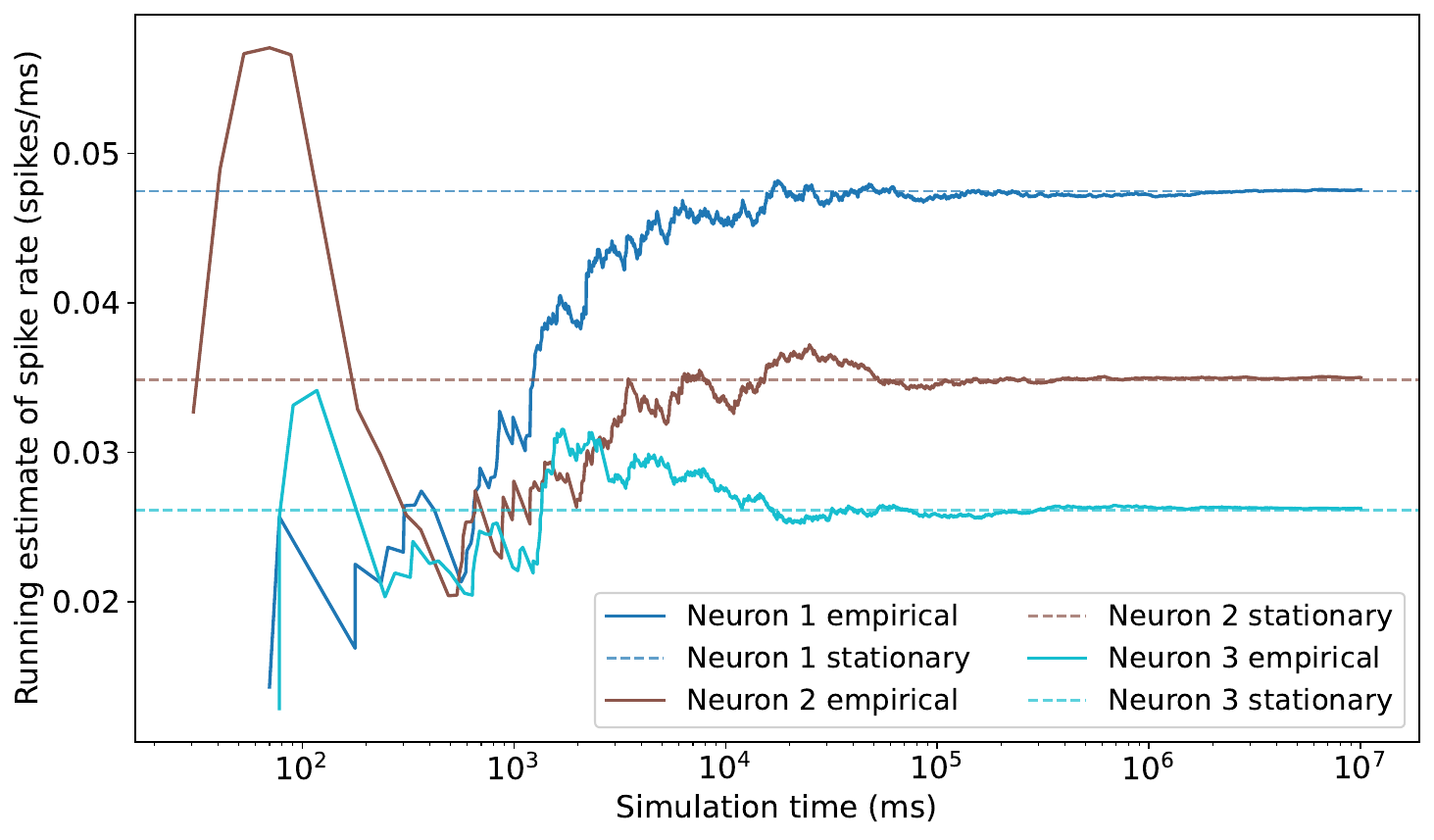}
  \caption{Convergence of firing rates in a three-neuron MM-RNN simulation.}
  \label{fig:convergence}
\end{figure}

As an initial validation, we compared simulation results against the analytical product-form solution of the standard synchronized RNN, replacing each neuron's static service rate with the stationary average firing rate of its Markov-modulated queue~\cite{gelenbe2008random}.
Under this approximation, simulated quantities such as the mean firing rate converge to the values predicted by the product-form formulas, as shown in Figure~\ref{fig:convergence}.
This suggests that the network may retain tractable stationary behavior under this approximation. 
The stationary behavior of the MM-RNN remains to be characterized. In particular, it is not yet established whether the product-form stationary distribution of the standard RNN is preserved under Markov-modulated service rates, or under what conditions the network remains stable.

This prototype provides a first building block toward a brain digital twin.
We plan to extend this approach to larger neuronal populations at the scale of a simulatable digital twin of brain electrical activity.
 \section{Conclusion}
No single model currently captures the mechanisms of both the single-neuron and population levels with fidelity. This limitation must be addressed to build brain digital twins. We proposed an MM-RNN model mapped onto a DEVS simulation infrastructure. The model is specified by a few high-level parameters. From these high-level parameters, technical parameters inherited from both parts of the model are generated automatically. For example, firing-rate regimes come from the MM component, while connectivity and population synchrony come from the RNN component. A DEVS simulation infrastructure is then produced. An analytical product-form approximation allows us to predict and validate stationary firing rates. This provides a first concrete step toward faithfully twinning brain electrical activity. Future work will focus on scaling the approach toward larger neuronal populations. In the longer term, such a model could contribute to brain digital twins for investigating pathological alterations in neuronal activity.


%
\bibliographystyle{splncs04}
\bibliography{refs}
\end{document}